\documentclass[a4paper,10pt,conference]{ieeeconf}
\IEEEoverridecommandlockouts                              
\usepackage{graphics} 
\usepackage{epsfig}   
\usepackage{mathptmx} 
\usepackage{times}    
\usepackage{amsmath}  
\usepackage{amssymb}  
\usepackage{txfonts}  

\title{\LARGE \bf
Error Analysis of Approximated PCRLBs for Nonlinear Dynamics}

\author{Ming Lei, Pierre Del Moral and Christophe Baehr %
\thanks{Dr. Ming Lei is with the INRIA BORDEAUX SUD-OUEST, University of Bordeaux-I in Bordeaux and CNRM/GAME URA1357, French National Centre for Meteorological Research in Toulouse, France. (\small minglei.sa@gmail.com)}%
\thanks{Prof. Pierre Del Moral is with the INRIA BORDEAUX SUD-OUEST and Mathematics Institute, University of Bordeaux-I, Bordeaux, France. (\small pierre.del\_moral@inria.fr)}%
\thanks{Prof. Christophe Baehr is with the CNRM/GAME URA1357, French National Centre for
        Meteorological Research, and the Toulouse Mathematics Institute, Statistic and Probability team, Toulouse, France. (\small christophe.baehr@math.univ-toulouse.fr)}%
}

\begin{document}
\maketitle
\thispagestyle{empty}
\pagestyle{empty}


\begin{abstract}
In practical nonlinear filtering, the assessment of achievable
filtering performance is important. In this paper, we focus on the
problem of efficiently approximate the posterior Cramer-Rao
lower bound (CRLB) in a recursive manner. By using Gaussian
assumptions, two types of approximations for calculating the CRLB
are proposed: An exact model using the state estimate as well as a Taylor-series-expanded model using both of the state estimate and its error covariance, are derived. Moreover, the difference between the two approximated CRLBs is also formulated analytically. By employing the particle filter (PF) and the unscented Kalman filter (UKF) to compute, simulation results reveal that the approximated CRLB using mean-covariance-based model outperforms that using the mean-based exact model. It is also shown that the theoretical difference between the estimated CRLBs can be improved through an improved filtering method.
\end{abstract}


\IEEEpeerreviewmaketitle

\section{Introduction}
\noindent It is well known that optimal estimators for the nonlinear
filtering of the discrete-time dynamic systems is an active area of
research and that a large number of suboptimal approximated
approaches were developed~\cite{Anderson79}. It is important to
quantify the accuracy of estimates obtained for the design of
algorithms such as the interacting multiple models (IMM) where
weighted estimates from multiple estimators are simultaneously
employed.

During the past thirty years many attempts have been made to
theoretically derive the achievable performance of nonlinear
filters. Deriving performance bounds are important since such bound
serve as indicators to measure system performance, and can be used
to determine whether imposed performance requirements are realistic
or not.

For dynamical statistical models, a commonly used bound is the CRLB
that has been investigated by various researchers: Van
Trees~\cite{Trees68} presented the batch form of a posterior CRLB
for random parameter vectors and a pre-1989 review~\cite{Kerr89}
summarized several lower bounds for nonlinear filtering, which
heavily emphasized the continuous time case.
Bobrovsky~\cite{Bobrovsky75} applied CRLB to discrete time problems
and Galdos~\cite{Galdos80} generalized it to the multi-dimensional
case. The main shortcoming of these formulations is the batch form
of implementation resulting high computational loads.
Tichavsky~\cite{Tichavsky98} was the first to derive a recursive
CRLB for updating the posterior Fisher information matrix (FIM) from
one time instance to the next while keeping the FIM constant in
size.

Subsequently, CRLB theory was extended to many applications, e.g.,
introducing the CRLB to multiple target tracking~\cite{Ristic04},
incorporating data association for tracking with the
CRLB~\cite{RuixinNiu01}, target detection for the case having a
detection probability less than unit~\cite{Farina02}, etc.

It is well known that the matrices in recursive form of FIM, can
only be theoretically determined by the true value of state.
Unfortunately, we cannot obtain the true state online in practice,
except in some well-designed experiments where true value of the
state is given as a prior knowledge. Therefore we naturally focus on
how to determine an approximate CRLB by using online
state estimates (as opposed to the true state values).

We have mainly two ways to approximate the CRLB \cite{Minglei}:
1) Make full use of the first-two order moments of the state estimate, i.e.,
expectation and covariance, by incorporating them with the Taylor
series expansion of the dynamics. 2) Combine the expectation of the
state with the exact dynamic model directly. The first method use
both estimates and is rather complex while the second method is
considerably simple, but depends heavily on an exact model. The
second method is mostly preferred in practice for its simpleness and
is sufficient to obtain an usable approximated CRLB.

The following question therefore needs to be addressed: By how much the
CRLB employed the two kinds of approximations differ from, and which one is
a better approximation to the true CRLB. This is the main motivation
of this investigation. In addition, determining the accuracy of the
estimated CRLB by using a state estimate, rather than the true state
under a recursive framework for a general nonlinear dynamics, has
not been addressed previously.

In this paper, we show how the state estimates can be applied to
determine the difference between the two estimated CRLBs. By using
Monte Carlo simulations, we show that the proposed method achieve a
satisfactory approximation, and the accuracy of estimated CRLB can
be explicitly improved by increasing the accuracy of filtering.

\section{Problem Formulation}
\subsection{Nonlinear Dynamical Model}
\noindent Consider the following discrete-time nonlinear dynamics with
additive Gaussian noise:
\begin{align}
&{\bf x}_{k+1} = {\bf f}_k({\bf x}_k) + {\bf w}_k,  \label{eq-1}\\
&{\bf z}_k = {\bf h}_k({\bf x}_k) + {\bf v}_k ,  \label{eq-2}
\end{align}
where the nonlinear vector-valued functions ${\bf f}_k \in
\mathbb{R}^{n \times 1}$ and ${\bf h}_k \in \mathbb{R}^{m \times 1}$
be used to model the state kinematics and measurement respectively, and generally $n > m$. ${\bf x}_k
\in \mathbb{R}^{n \times 1}$ is the state vector, ${\bf z}_k \in
\mathbb{R}^{m \times 1}$ is the measurement vector, ${\bf w}_k \in
\mathbb{R}^{n \times 1}$ is a zero-mean white Gaussian process noise
with known covariance ${\bf Q}_k$, and ${\bf v}_k \in \mathbb{R}^{m
\times 1}$ a zero-mean Gaussian white measurement noise with
variance ${\bf R}_k$. The initial state ${\bf x}_0$ is
assumed as a Gaussian distribution with mean ${\bar{\bf x}}_0$ and variance ${\bf P}_0$. Moreover, a general accepted assumption like $cov({\bf x}_0,{\bf v}_k)=0, cov({\bf x}_0,{\bf w}_k)=0$.

\subsection{Posterior CRLB }
\noindent Let ${\hat{\bf x}}_k$ and ${\bf C}_k$ denote the unbiased
state estimate and its error covariance at time instant
$k$. We therefore have
\begin{equation}\label{eq-3}
   {\bf C}_k = E\!\left[{\tilde{\bf x}}_k{\tilde{\bf x}}'_k\right] \geq {\bf
   J}_k^{-1} ,
\end{equation}
where ${\tilde{\bf x}}_k = {\bf x}_k - {\hat{\bf x}}_k$ is
the prediction error of state. ${\bf J}_k^{-1}$ is the posterior
CRLB (PCRLB), defined to be the inverse of FIM, ${\bf J}_k$. The superscript $(\cdot)'$ in~\eqref{eq-3}
denotes the transpose of a vector or a matrix, and the inequality
in~\eqref{eq-3} means that the difference ${\bf C}_k - {\bf
J}_k^{-1}$ is a positive semidefinite matrix.
From~\cite{Tichavsky98, Simandl01} we know that the sequential FIM ${\bf J}_k$ can be recursively calculated by
\begin{align}
    &{\bf J}_{k+1} = {\bf D}_k^{22} - {\bf D}_k^{21}({\bf J}_k +
    {\bf D}_k^{11})^{-1}{\bf D}_k^{12} \quad (k > 0), \label{eq-4}\\
    &{\bf J}_0 = E\left[-\Delta_{{\bf x}_0}^{{\bf x}_0}\log p({\bf x}_0)\right], \label{eq-5} \\
    &{\bf D}_k^{11} = E\left[-\Delta_{{\bf x}_k}^{{\bf x}_k}\log p({\bf x}_{k+1}|{\bf x}_k)\right],\label{eq-6}\\
    &{\bf D}_k^{12} = \left({\bf D}_k^{21}\right)' = E\left[-\Delta_{{\bf x}_k}^{{\bf x}_{k+1}}\log p({\bf x}_{k+1}|{\bf x}_k)\right],\label{eq-7}\\
\nonumber
    &{\bf D}_k^{22} = E\left[-\Delta_{{\bf x}_{k+1}}^{{\bf x}_{k+1}}\log p({\bf x}_{k+1}|{\bf x}_k)\right]+ \\
    &\quad \quad E\left[-\Delta_{{\bf x}_{k+1}}^{{\bf x}_{k+1}}\log p({\bf z}_{k+1}|{\bf x}_{k+1})\right], \label{eq-8}
\end{align}
here let $\nabla$ and $\Delta$ be operators of the first and
second-order partial derivatives, i.e., $\nabla_{\bf x} = \left[ \frac{\partial}{\partial x_1},\cdots,\frac{\partial}{\partial x_n}\right]', \Delta_{\bf x}^{\bf y} = \nabla_{\bf x}(\nabla_{\bf y})'$.
Note that all the above expectations are taken with respect to the
joint probability density function (PDF) $p({\bf x}_{0:k+1}|{\bf
z}_{1:k+1})$, where ${\bf x}_{0:k+1}$ and ${\bf z}_{1:k+1}$ denote
all the states and measurements up to time $k+1$.

\section{Approximated Gaussian Form (AGF) of Nonlinear Dynamics}

\noindent According to CRLB theory, the derivatives in \eqref{eq-4} should be
evaluated at the true value of state ${\bf x}_k$. Our final aim is to use
the moments of state estimate instead of the true state to
calculate the difference between the approximated PCRLBs, thus the FIM matrices (i.e., ${\bf D}_k^{11}$, ${\bf D}_k^{12}$ and ${\bf D}_k^{22}$ should be represented, therefore, the density function $p({\bf x}_{k+1}|{\bf x}_k)$ and $p({\bf z}_{k+1}|{\bf x}_{k+1})$ from \eqref{eq-1} and \eqref{eq-2}
 should be firstly formulated explicitly.

\subsection{AGF by the First-two Order Moment of State Estimate}\label{approxII}

\noindent
Assume that the first and second moment estimation of state ${\bf
x}_k$ is known and given by ${\hat{\bf x}}_k$ and ${\hat{\bf
P}_k^{\bf f}}=E\left[{\tilde{\bf x}}_k{\tilde{\bf x}}'_k|{\bf z}_{1:k}\right]$, and also assume that the distribution of ${\bf x}_{k+1}$ can be approximated by a Gaussian. We immediately have
\begin{equation}\label{eq-9}
    {\bf x}_{k+1} \approx \mathcal{N}\!\left[{\bf x}_{k+1};~{\bar{\bf x}}_{k+1},~{\bf P}_{k+1}^{\bf
    x}\right] ,
\end{equation}
where ${\bar {\bf x}}_{k+1}=E\left[{\bf x}_{k+1}|{\bf z}_{1:k}\right]\approx
{\hat{\bf f}}_k + {\breve{\bf f}}_k$, in which ${\hat{\bf f}}_k = {\bf
f}_k({\hat{\bf x}}_k)$, ${\breve{\bf f}}_k
=\frac{1}{2}\sum_{i=1}^n{\bf e}_i tr\left[{\hat{\bf S}}_{k,i}^{{\bf
f}}{\hat{\bf P}}_k^{{\bf f}}\right]$, ${\bf e}_i \in \mathbb{R}^{n \times 1}$
denotes the $i$-th unit normal vector in column shape, and
$tr[\cdot]$ denotes trace operation. ${\hat{\bf S}}_{k,i}^{\bf
f}=\nabla _{{\bf x}_k}\left[\nabla _{{\bf x}_k}f_{k,i}({\bf x}_k)\right]'$ is
the Hessian matrix of $i$-th element $f_{k,i}({\bf x}_k)$ of the
vector-valued function ${\bf f}_k({\bf x}_k)$. Notation ${\hat{\bf
F}}_k^{\bf f}=\left[\nabla_{{\bf x}_k}{\bf f}'_k({\bf x}_k)\right]'=\left[\partial
f_{k,i}({\bf x}_k)/\partial x_j\right]_{n \times n}$ denotes the Jacobian
matrix with $n \times n $ dimension, ${\bf P}_{k+1}^{\bf x} =
{\breve{\bf P}}_{k+1}^{\bf x} + {\bf Q}_k$, ${\breve{\bf
P}}_{k+1}^{\bf x} = {\hat{\bf F}}_k^{\bf f}{\hat{\bf P}}_k^{\bf
f}({\hat{\bf F}}_k^{\bf f})' + \frac{1}{2}\sum_{i=1}^n \sum_{j=1}^n {\bf e}_i{\bf e}'_j tr\left[{\hat{\bf S}}_{k,i}^{\bf f}{\hat{\bf
P}}_k^{\bf f}{\hat{\bf S}}_{k,j}^{\bf f}{\hat{\bf P}}_k^{\bf f}\right]$.
Similar to \eqref{eq-9}, the Gaussian form of the measurement ${\bf
z}_k$ can be approximated by
\begin{equation}\label{eq-10}
    {\bf z}_k \approx \mathcal{N}\left[{\bf z}_k;~{\bar{\bf z}}_k,~{\bf P}_k^{\bf z}\right] ,
\end{equation}
where the expectation ${\bar{\bf z}}_k = E\left[{\bf z}_k|{\bf
x}_k\right]\approx {\hat{\bf h}}_k + {\breve{\bf h}}_k$, the covariance
${\bf P}_k^{\bf z} \approx {\bf R}_k + {\breve{\bf P}}_k^{\bf z}$,
in which ${\breve{\bf h}}_k = \frac{1}{2}\sum_{i=1}^m {\bf e}_i
tr\left[{\hat{\bf S}}_{k,i}^{\bf h}{\hat{\bf P}}_k^{\bf h}\right]$,
${\breve{\bf P}}_k^{\bf z} = {\hat{\bf F}}_k^{\bf h}{\hat{\bf
P}}_k^{\bf h}({\hat{\bf F}}_k^{\bf h})' + \frac{1}{2}\sum_{i=1}^m
\sum_{j=1}^m {\bf e}_i {\bf e}'_j tr\left[{\hat{\bf S}}_{k,i}^{\bf
h}{\hat{\bf P}}_k^{\bf h}{\hat{\bf S}}_{k,j}^{\bf h}{\hat{\bf
P}}_k^{\bf h}\right]$.  The terms ${\hat{\bf h}}_k$, ${\hat{\bf F}}_k^{\bf
h}$ and ${\hat{\bf S}}_{k,i}^{\bf h}$ are similar to the definitions
of ${\hat{\bf f}}_k$, ${\hat{\bf F}}_k^{\bf f}$ and ${\hat{\bf
S}}_{k,i}^{\bf f}$ in \eqref{eq-9}, respectively.

\subsection{AGF by the First Order Moment of State Estimate}
\noindent
As an alternative to the approximation presented in Section
\ref{approxII}, we use the state estimate ${\hat{\bf x}}_k$ to
represent ${\bf x}_{k+1}$ and ${\bf z}_k$. By denoting this version of
representation as ${\bf x}_{k+1}^{\ast}$ and ${\bf z}_k^{\ast}$, we have
\begin{align}
    &{\bf x}_{k+1}^{\ast} \approx \mathcal{N}\!\left[{\bf x}_{k+1}^{\ast};
    ~{\hat{\bf f}}_k,~{\bf Q}_k\right], \label{eq-11} \\
    &{\bf z}_k^{\ast} \approx \mathcal{N}\!\left[{\bf z}_k^{\ast};
    ~{\hat{\bf h}}_k,~{\bf R}_k\right], \label{eq-12}
\end{align}
where the definitions of ${\hat{\bf f}}_k$ and ${\hat{\bf h}}_k$ are
same as that in Section \ref{approxII}.

\section{Approximated FIM}
\subsection{The Case Using Mean and Covariance }\label{mean+cov}
\noindent According to distribution of ${\bf x}_{k+1}$ and ${\bf
z}_k$ in \eqref{eq-9} and \eqref{eq-10}, the log-PDF of state and
measurement, given by ${\bf x}_k$ and ${\bf x}_{k+1}$, can be
respectively formulated by
\begin{align}
\nonumber
    &\log p({\bf x}_{k+1}|{\bf x}_k) = c_1 - \frac{1}{2}\log\det\left[{\bf P}_{k+1}^{\bf x}\right] -\frac{1}{2}\left[({\bf x}_{k+1}-{\bar{\bf x}}_{k+1})' \right. \\
    & \quad \quad \left. \times ({\bf P}_{k+1}^{\bf x})'({\bf x}_{k+1}-{\bar{\bf x}}_{k+1})\right], \label{eq-13}\\
\nonumber
    &\log p({\bf z}_{k+1}|{\bf x}_{k+1}) = c_2 - \frac{1}{2}\log\det\left[{\bf P}_{k+1}^{\bf z}\right] -\frac{1}{2}\left[({\bf z}_{k+1}-{\bar{\bf z}}_{k+1})' \right. \\
    & \quad \quad \left. \times ({\bf P}_{k+1}^{\bf z})'({\bf z}_{k+1}-{\bar{\bf z}}_{k+1})\right], \label{eq-14}
\end{align}
where $c_1$ and $c_2$ are constants. Calculate the derivatives of $\log p({\bf x}_{k+1}|{\bf
x}_k)$ and $\log p({\bf z}_{k+1}|{\bf x}_{k+1})$ with respective to
${\bf x}_k$ and ${\bf x}_{k+1}$ respectively, specifically we have
\begin{equation}
    \nabla_{{\bf x}_{k+1}}\left[\log p({\bf x}_{k+1}|{\bf x}_k)\right] =
    -({\bf P}_{k+1}^{\bf x})^{-1}({\bf x}_{k+1}- {\bar{\bf
    x}}_{k+1}) , \label{eq-15}
\end{equation}
then consider the definitions of FIM in \eqref{eq-6}-\eqref{eq-8} and after algebra arrangement, finally we obtain
\begin{align}
\nonumber
    {\bf D}_k^{11} =& \sum_{i=1}^n\sum_{j=1}^n{\bf e}_i{\bf
    e}'_j \left[\frac{\partial{\bar{\bf x}}'_{k+1}}{\partial
    x_k^i}({\bf P}_{k+1}^{\bf x})^{-1}\frac{\partial{\bar{\bf x}}_{k+1}}{\partial x_k^j} \right. + \\
    & \left. \frac{1}{2}tr\left(({\bf P}_{k+1}^{\bf x})^{-1}
    \frac{\partial{\bf P}_{k+1}^{\bf x}}{\partial x_k^i}({\bf P}_{k+1}^{\bf x})^{-1}\frac{\partial{\bf P}_{k+1}^{\bf x}}{\partial x_k^j} \right) \right], \label{eq-16}\\
    {\bf D}_k^{12} =& -\frac{\partial{\bar{\bf x}}'_{k+1}}{\partial{\bf
    x}_k}({\bf P}_{k+1}^{\bf x})^{-1},  \label{eq-17}\\
\nonumber
    {\bf D}_k^{22} =& ({\bf P}_{k+1}^{\bf x})^{-1} \!+\! \sum_{i=1}^n\sum_{j=1}^n{\bf e}_i{\bf
    e}'_j \! \left[\frac{\partial{\bar{\bf z}}'_{k+1}}{\partial
    x_{k+1}^i}({\bf P}_{k+1}^{\bf z})^{-1}\frac{\partial{\bar{\bf z}}_{k+1}}{\partial x_{k+1}^j} \right. \\
    & + \! \left. \frac{1}{2}tr\left(({\bf P}_{k+1}^{\bf z})^{-1}
    \frac{\partial{\bf P}_{k+1}^{\bf z}}{\partial x_{k+1}^i}({\bf P}_{k+1}^{\bf z})^{-1}
    \frac{\partial{\bf P}_{k+1}^{\bf z}}{\partial x_{k+1}^j} \right) \right]. \label{eq-18}
\end{align}
It is explicit that the right hand of \eqref{eq-16} and the second term on the right hand of \eqref{eq-18} is similar with that in~\cite{Key93}. We observe that all derivatives involved in \eqref{eq-16}-\eqref{eq-18} can be evaluated by using the mean and covariance of the state estimate instead of the true state.

So far, based on the Gaussian model assumption, we formulate the
matrices used by the PCRLB in \eqref{eq-4} as above. In order to obtain the
difference between the two kinds of approximated PCRLBs, matrices
in \eqref{eq-16}-\eqref{eq-18} should be decomposed as shown in the
follows. According to the well-known matrix inversion lemma~\cite{Eves66}, we have a simplified formulas as below
\begin{equation}\label{eq-19}
    ({\bf A} + {\bf B})^{-1} = {\bf A}^{-1} -
    ({\bf A}{\bf B}^{-1}{\bf A} + {\bf A})^{-1},
\end{equation}
where ${\bf A}$, ${\bf B}$ are the nonsingular matrices, and the
inversion of every matrix is assumed to exist. For the matrix ${\bf
D}_k^{11}$, we can decompose the inversion of the covariance matrix
${\bf P}_{k+1}^{\bf x}$ defined in \eqref{eq-9} into two terms,
$({\bf P}_{k+1}^{\bf x})^{-1}={\bf Q}_k^{-1} - {\bf \Psi}_k^{\bf
x}$, where ${\bf \Psi}_k^{\bf x} = \left[{\bf Q}_k ({\breve{\bf
P}}_k^{\bf x})^{-1}{\bf Q}_k + {\bf Q}_k\right]^{-1}$. Substituting it and
the expression of ${\bar{\bf x}}_{k+1}$ into \eqref{eq-16}, after
some arrangements yield
\begin{equation}\label{eq-20}
    {\bf D}_k^{11} = \underbrace{\sum_{i=1}^n\sum_{j=1}^n{\bf e}_i{\bf
    e}'_j \left(\frac{\partial {\hat{\bf f}'_k}}{\partial x_k^i}{\bf Q}_k^{-1}\frac{\partial {\hat{\bf f}_k}}{\partial x_k^j} \right)}_{\Sigma_{11}^{\ast}} + \Sigma_{11} ,
\end{equation}
where
\begin{align*}
    &\Sigma_{11} = \frac{1}{2}\sum_{i=1}^n\sum_{j=1}^n{\bf e}_i{\bf e}'_j
    tr \left(({\bf P}_{k+1}^{\bf x})^{-1} \frac{\partial{{\bf P}_{k+1}^{\bf x}}}{\partial x_k^i} ({\bf P}_{k+1}^{\bf x})^{-1} \frac{\partial {{\bf P}_{k+1}^{\bf x}}} {\partial x_k^j} \right)\\
    & + \! \sum_{i=1}^n\sum_{j=1}^n{\bf e}_i{\bf e}'_j \left[ \frac{\partial {\breve{\bf f}'_k}}{\partial x_k^i}{\bf Q}_k^{-1}
    \frac{\partial {\hat{\bf f}_k}}{\partial x_k^j}
     + \frac{\partial {\bar{\bf x}'_{k+1}}}{\partial x_k^i}
    \left({\bf Q}_k^{-1} \frac{\partial {\breve{\bf f}_k}}{\partial x_k^j} - {\bf \Psi}_k^{\bf x} \frac{\partial {\bar{\bf x}_{k+1}}}{\partial x_k^j} \!\right)\right].
\end{align*}
For matrix ${\bf D}_k^{22}$, we decompose $({\bf P}_{k+1}^{\bf
z})^{-1} = {\bf R}_{k+1}^{-1} - {\bf \Psi}_{k+1}^{\bf z}$, where
${\bf \Psi}_{k+1}^{\bf z} = \left[{\bf R}_{k+1}({\breve{\bf
P}}_{k+1}^{\bf z})^{-1}{\bf R}_{k+1} + {\bf R}_{k+1}\right]^{-1}$.
Substituting it and the expression of ${\bar{\bf z}}_k$ into
\eqref{eq-18} yields
\begin{equation}\label{eq-21}
    {\bf D}_k^{22} = \underbrace{{\bf Q}_k^{-1} + \sum_{i=1}^n\sum_{j=1}^n{\bf e}_i{\bf e}'_j \left( \frac{\partial {\hat{\bf h}'_{k+1}}}{\partial x_{k+1}^i}{\bf R}_{k+1}^{-1} \frac{\partial {\hat{\bf h}_{k+1}}}{\partial x_{k+1}^j} \right)}_{\Sigma_{22}^{\ast}} + \Sigma_{22} ,
\end{equation}
where
\begin{align*}
    \Sigma_{22} =& \frac{1}{2}\sum_{i=1}^n\sum_{j=1}^n{\bf e}_i{\bf e}'_j
    tr \left(({\bf P}_{k+1}^{\bf z})^{-1} \frac{\partial{{\bf P}_{k+1}^{\bf z}}}{\partial x_{k+1}^i} ({\bf P}_{k+1}^{\bf z})^{-1} \frac{\partial {{\bf P}_{k+1}^{\bf z}}} {\partial x_{k+1}^j} \right) \\
    & - \! {\bf \Psi}_k^{\bf x} + \! \sum_{i=1}^n\sum_{j=1}^n{\bf e}_i{\bf e}'_j \left[ \frac{\partial {\breve{\bf h}'_{k+1}}}{\partial x_{k+1}^i}{\bf R}_{k+1}^{-1} \frac{\partial {\hat{\bf h}_{k+1}}}{\partial x_{k+1}^j} \right. \\
    & \left. + \frac{\partial {\bar{\bf z}'_{k+1}}}{\partial x_{k+1}^i}
    \! \left({\bf R}_{k+1}^{-1} \frac{\partial {\breve{\bf h}'_{k+1}}}{\partial x_{k+1}^j} - {\bf \Psi}_{k+1}^{\bf z} \frac{\partial {\bar{\bf z}_{k+1}}}{\partial x_{k+1}^j} \right)\right].
\end{align*}
For matrix ${\bf D}_k^{12}$, substituting $({\bf P}_{k+1}^{\bf
x})^{-1} = {\bf Q}_k^{-1} - {\bf{\Psi}}_k^{\bf x}$ and ${\bar{\bf
x}}_{k+1} \approx {\hat{\bf f}}_k + {\breve{\bf f}}_k$ into
\eqref{eq-17} yields
\begin{equation}\label{eq-22}
    {\bf D}_k^{12} = \underbrace{-\frac{\partial{\hat{\bf f}}'_k}{\partial{\bf     x}_k}{\bf Q}_k^{-1}}_{\Sigma_{12}^{\ast}} + \underbrace{\left( \frac{\partial {\hat{\bf f}}'_k}{\partial{\bf x}_k} + \frac{\partial {\breve{\bf f}}'_k}{\partial{\bf x}_k} \right)\!{\bf \Psi}_k^{\bf x} - \frac{\partial {\breve{\bf f}}'_k}{\partial{\bf
    x}_k}{\bf Q}_k^{-1}}_{\Sigma_{12}} .
\end{equation}
So after the above steps, we successfully rewrite the matrices ${\bf D}_k^{11}$, ${\bf D}_k^{22}$ and ${\bf D}_k^{12}$ into two parts respectively, then we submit expressions in \eqref{eq-20}-\eqref{eq-22} into the definition of FIM in \eqref{eq-4}, using the matrix inversion lemma again, after some expansions and arrangements yield
\begin{equation}\label{eq-23}
    {\bf J}_{k+1} = \underbrace{\Sigma_{22}^{\ast} - \Sigma_{12}'^{\ast}\left({\bf J}_k + \Sigma_{11}^{\ast}\right)^{-1}\Sigma_{12}^{\ast}}_{\bf \Theta} + {\bf
    \Pi} ,
\end{equation}
where
\begin{align*}
    &{\bf \Pi} = \Sigma_{22} - ({\bf D}_k^{12})'({\bf J}_k + {\bf
    D}_k^{11})^{-1}\Sigma_{12} - \\
    & \quad \quad \left[\Sigma'_{12}({\bf J}_k + {\bf D}_k^{11})^{-1} - \Sigma_{12}'^{\ast}{\bf \Phi} \right] \Sigma_{12}^{\ast} , \\
    &{\bf \Phi} = \left[({\bf J}_k + \Sigma_{11}^{\ast})(\Sigma_{11})^{-1}
    ({\bf J}_k + \Sigma_{11}^{\ast}) + ({\bf J}_k +
    \Sigma_{11}^{\ast})\right]^{-1} .
\end{align*}

\subsection{The Case Using only Mean}

\noindent By comparing the mean-based Gaussian form presented in
\eqref{eq-16}-\eqref{eq-18}, we straightforwardly arrive at (use the
superscript symbol $\ast$ to distinguish with that in Section
\ref{mean+cov}): ${\bf D}_k^{11\ast} = \Sigma_{11}^{\ast}$, ${\bf
D}_k^{12\ast} = \Sigma_{12}^{\ast}$ and ${\bf D}_{k+1}^{22\ast} =
\Sigma_{22}^{\ast}$. Then substituting matrices of ${\bf
D}^{\ast}$s into the definition of FIM in~\eqref{eq-4} yields
\begin{equation}\label{eq-24}
    {\bf J}_{k+1}^{\ast} = {\bf D}_{k+1}^{22\ast} -
     ({\bf D}_k^{12\ast})'({\bf J}_k + {\bf D}_k^{11\ast})^{-1}{\bf
     D}_k^{12\ast} \equiv {\bf \Theta} .
\end{equation}

\section{Difference Between the Two PCRLB Approximations}

\noindent Our final aim is to calculate the difference between the two approximated PCRLBs, where the one approximation employ the first-two order moment of state estimate and the other one only use the first order
moment.

Performing the matrix inversion lemma on the FIM ${\bf J}_{k+1}$ defined in \eqref{eq-23} again, we get the PCRLB ${\bf J}_{k+1}^{-1}$ directly
\begin{equation}\label{eq-25}
    {\bf J}_{k+1}^{-1} = {\bf \Theta}^{-1} -
    ({\bf \Pi}^{-1}{\bf \Theta} + {\bf I})^{-1}{\bf \Theta}^{-1} .
\end{equation}
Explicitly the difference between the two kinds of approximated PCRLBs,
defined by ${\tilde{\bf J}}_{k+1}^{-1} \triangleq {\bf J}_{k+1}^{*-1}
- {\bf J}_{k+1}^{-1}$, can be formulated by
\begin{equation} \label{eq-26}
    {\tilde{\bf J}}_{k+1}^{-1} =
    ({\bf \Pi}^{-1}{\bf \Theta} + {\bf I})^{-1}{\bf \Theta}^{-1}
    = ({\bf \Pi}^{-1}{\bf J}_{k+1}^{\ast} + {\bf I})^{-1}({\bf
    J}_{k+1}^{\ast})^{-1},
\end{equation}
where ${\bf I}$ denotes an identity matrix with appropriate
dimension. In Section \ref{simu}, Monte Carlo simulations show that
the bound ${\bf J}_{k+1}^{\ast-1}$ is always higher than the ${\bf
J}_{k+1}^{-1}$, that is to say, the ${\bf J}_{k+1}^{-1}$ is more closer to
the true PCRLB than that of ${\bf J}_{k+1}^{\ast-1}$, of course this is for the case
with finite number of particles. For the situation as sampling $N$ tends to infinity,
the convergence theoretically needs further investigation.

\section{Experimental Results}\label{simu}
\noindent To evaluate the performance of the proposed algorithm, the
following typical univariate nonlinear model~\cite{Kotecha03} is
studied:
\begin{equation} \label{eq-27}
\left\{ \begin{aligned}
    x_k =& 0.5 x_{k-1} \!+ \!\frac{25x_{k-1}}{1+x_{k-1}^2}
    + 8\cos[1.2(k-1)] + w_k , \\
    y_k =& \frac{x_k^2}{20} + v_k, \quad k=1,2,\cdots,T
    \end{aligned}
\right.
\end{equation}
here using $w_k\sim \mathcal{N}(0,\sigma_w^2)$ denotes the process
noise, and $v_k\sim \mathcal{N}(0,\sigma_v^2)$ is the measurement
noise. Data was generated by using $\sigma_w^2=1$, $\sigma_v^2=5$,
and $T=50$. The initial prior distribution was chosen as $p(x_0)\sim
20\times \mathcal{N}(0,1)$.
\begin{figure}
\begin{center}
\includegraphics[width=3.3in]{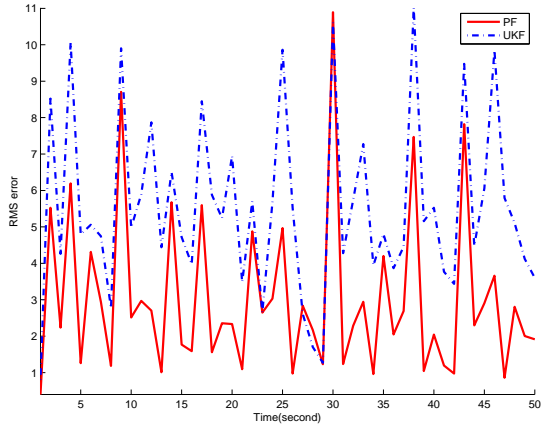}
\end{center}
\caption{Comparison RMS errors of state estimation generated by two
different estimators: Particle filter (PF) and Unscented Kalman
filter (UKF). 100 runs of Monte Carlo simulations and the initial
number of particles is $1000$. }\label{fig1}
\end{figure}

\begin{figure}[h]
\begin{center}\leavevmode
\includegraphics[width=3.3in]{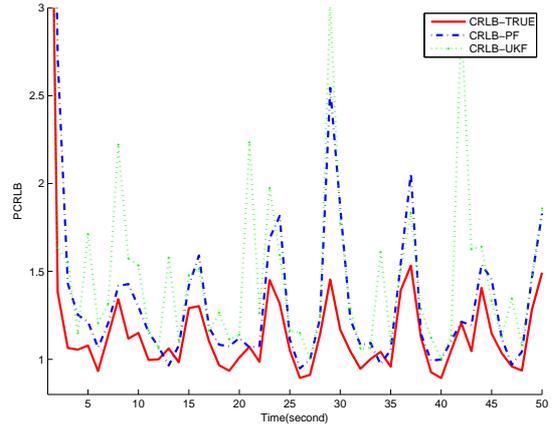}
\end{center}
\caption{Comparison of the true posterior CRLB with the first type
of approximations. The approximated PCRLB corresponds to the method
of ``exact model and expectation of state estimation", and were
generated by Particle filter (PF) and Unscented Kalman filter (UKF).
100 runs of Monte Carlo simulation. }\label{fig2}
\end{figure}

\begin{figure}[h]
\begin{center}\leavevmode
\includegraphics[width=3.3in]{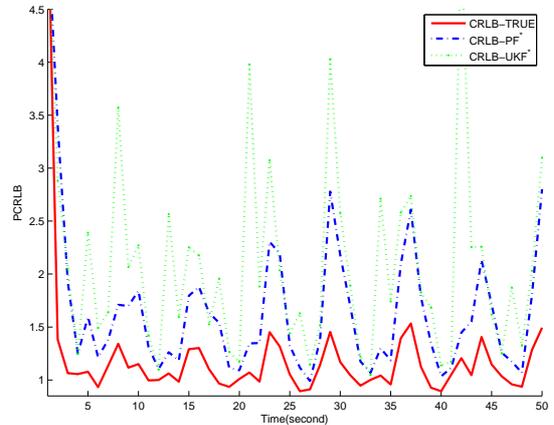}
\end{center}
\caption{Comparison of the true posterior CRLB with the second type
of approximations. The approximated PCRLB corresponds to the method
of ``Taylor expanded model and first-two moments of state
estimation", the two estimators: Particle filter (PF) and Unscented
Kalman filter (UKF), were employed. $100$ runs of Monte Carlo
simulation. }\label{fig3}
\end{figure}

\begin{figure}[h]
\begin{center}\leavevmode
\includegraphics[width=3.3in]{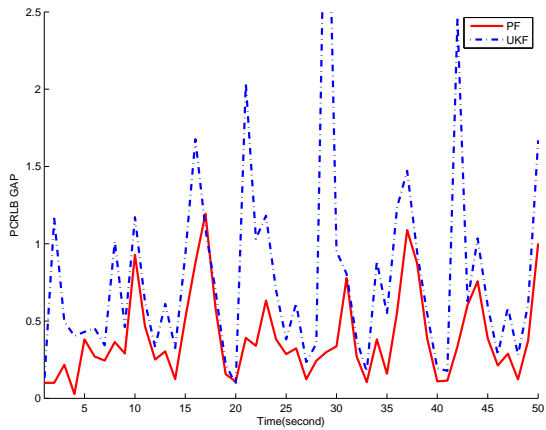}
\end{center}
\caption{Comparison of two kinds of theoretical gap of posterior
CRLB, upper and lower plots corresponds to Unscented Kalman filter
(UKF) and Particle filter (PF), respectively. As expected, the gap
improve as the accuracy of filtering improves. }\label{fig4}
\end{figure}

For comparison purposes, we implemented two state estimation
methods: 1) The unscented Kalman filter (UKF) where it is not
necessary to compute Jacobian matrices and the performance is
accurate to the third-order term (in the Taylor series expansion)
for Gaussian inputs, even for nonlinear systems. For non-Gaussian
inputs, approximations are accurate to at least the second-order
term~\cite{Julier95}. 2) The particle filter (PF), where an initial
sample size $N=1000$ is adopted, and $100$ runs of Monte Carlo
simulation are performed.

Filtering accuracy by using the same trajectories is shown in Fig.1.
Here the root mean square (RMS) error is used as an evaluation
criterion. It should be firstly noted that for the PF the initial
number of samples is generally chosen by trial-and-error and that
its accuracy can be improved by increasing the sample size.
Secondly, according to~\cite{Kotecha03}, the likelihood $p(y_k|x_k)$
has a bimodal nature when $x_k>0$, and this bimodality causes the
state too acutely fluctuate and complicates to track using
conventional filtering. The RMS error in Fig.\ref{fig1} clearly
reflects the effect of the nonlinear dynamic phenomena.

Fig.\ref{fig2} shows the comparison of the true PCRLB and
the approach of ``exact model and mean of state estimation", which
refers the recursive FIM formulated by \eqref{eq-24}. We can see
from the figure that there exists an explicit error between the true
PCRLB and both approximations. The PCRLB corresponding to UKF is
overall worse than the PCRLB generated by the PF. As expected, the
true PCRLB is a lower bound (always lower than the approximations in
all instants).

In Fig.\ref{fig3}, the true posterior CRLB is compared with the
approach of ``Taylor series expanded model and first-two order
moments of state estimation". This approach is performed by
substituting Eqn.\eqref{eq-16}-\eqref{eq-18} into \eqref{eq-4} and
using first-two order moments of state estimation as parameters.
we observe that both estimated PCRLBs are more accurate approximations 
compared with the true PCRLB. In many instants the PCRLB corresponding 
to PF is closer to the true PCRLB than the approximation using the UKF. 
Due to the acute nonlinearity of the system, the PCRLBs appear strongly 
oscillatory throughout the simulation.

We can directly calculate the difference between the two PCRLB
approximations: PCRLB in Fig.\ref{fig2} minus the corresponding one
in Fig.\ref{fig3}. However, as a theoretical analysis, we employ the
formula in \eqref{eq-26} and the calculated results are presented in
Fig.\ref{fig4}. The PCRLB generated by the PF is generally more
accurate throughout simulation. When the initial sampling used by PF
was increased, the accuracy of its corresponding PCRLB was improved.

\section{Conclusion}
\noindent 
In this paper, we considered the problem of approximate calculation 
of CRLB by using Gaussian assumptions and the moments of state estimate
instead of using true state. Two kinds of approaches were proposed:
One was an exact model using the expectation of state estimate; the
other was an approximated model using the expectation and covariance
of state estimate. Furthermore, the difference between the two
estimated CRLBs was formulated analytically. By using state estimators 
of PF and UKF, we compared the proposed approximations with true PCRLB.
Simulation results demonstrated the significance and validity of our
approach.



\end{document}